\documentclass[%
 reprint,
superscriptaddress,
 amsmath,amssymb,
 aps,
]{revtex4-2}

\usepackage{graphicx}% Include figure files
\usepackage{dcolumn}% Align table columns on decimal point
\usepackage{bm}% bold math
\usepackage{xcolor,colortbl}
\usepackage{float}
\usepackage[colorlinks=true, allcolors=blue]{hyperref}
\usepackage[colorinlistoftodos,textsize=tiny]{todonotes}
\usepackage{mathtools} 
\graphicspath{ {./images/} }
\begin{document}

\preprint{APS/123-QED}
\title{Non-equilibrium fluctuations and nonlinear response of an active bath}

\author{Hunter Seyforth}
\thanks{equally contributing authors}
\affiliation{Department of Physics,  California State University Fullerton, CA 92831 USA}

\author{Mauricio Gomez}
\thanks{equally contributing authors}
\affiliation{Department of Physics,  California State University Fullerton, CA 92831 USA}

\author{W. Benjamin Rogers}
\affiliation{Department of Physics,  Brandeis University, Waltham, MA 02452 USA}

\author{Jennifer L. Ross}
\affiliation{Department of Physics,  Syracuse University, Syracuse, NY 13244 USA}

\author{Wylie W. Ahmed}
\email[correspondence: ]{wahmed@fullerton.edu }
\affiliation{Department of Physics,  California State University Fullerton, CA 92831 USA}


\date{\today}% It is always \today, today,

\begin{abstract}
We analyze the dynamics of a passive colloidal probe immersed in an active bath  using an optical trap to study three physical processes: (1) the non-equilibrium fluctuations transferred to the probe by the active bath, (2) the friction experienced by the probe as it is driven through the active bath, and (3) the force relaxation of the probe returning to its equilibrium position. We measure the local force dynamics where all of the following characteristics are of $\mathcal{O}(1)$: the size of the probe colloid relative to active bath particle; the size of the probe colloid relative to the characteristic run-length of an active particle; and the timescale of probe movement to the persistence time of an active particle.  We find at P\'{e}clet (Pe) $\ll 1$ the active suspension exhibits shear thinning down to the solvent viscosity (but not below); at $0.85 <$ Pe $\leq 5.1$ the active bath shear thickens; and at Pe $\geq 8.5$ the effective viscosity of the active bath shows a decreased effect of thickening and plateaus.  These results are in agreement with recent modeling and simulations of the nonlinear rheology of an isotropic active bath, providing experimental verification, and suggesting the model predictions extends to moderately dense suspensions.  Further, we observe that the distribution of force fluctuations depends on Pe, unlike in passive equilibrium baths.  Lastly, we measure the energy transfer rate from the active bath to the probe to be $\langle J \rangle \approx 10^3$ $k_B T/$s, which leads to an increase in the effective diffusion of the probe by a factor of $\sim 2$.
\end{abstract}

\maketitle

%\tableofcontents

\maketitle

\section{Introduction}

Collections of self-propelled particles have become a cornerstone for theoretical and experimental studies of active matter~\cite{bechinger2016active, popkin2016physics, fodor2018statistical}.  Model systems (living and non-living) cover a wide range of length scales, from nanometer to meter~\cite{ghosh2021enzymes,  elgeti2015physics, alert2021living, wensink2012meso, bricard2013emergence, giomi2013swarming, scholz2018inertial,gouiller2021mixing, cavagna2014bird, ginelli2015intermittent}, but they all share a common trait: the individual objects that compose the system consume energy and generate self-propulsion~\cite{ramaswamy2010mechanics,marchetti2013hydrodynamics, bechinger2016active}.  Consequently, these systems are far from equilibrium and exhibit interesting dynamics, such as violation of the fluctuation dissipation theorem (FDT)~\cite{chen2007fluctuations, ahmed2015active, mizuno2007nonequilibrium, martin2001comparison}, broken detailed balance~\cite{battle2016broken, martinez2019inferring, gnesotto2018broken, tabatabai2021detailed}, entropy production~\cite{nardini2017entropy, dabelow2019irreversibility, shankar2018hidden, pigolotti2017generic},  collective motion~\cite{dombrowski2004self, vicsek2012collective, sokolov2012physical, wioland2016directed}, giant density fluctuations~\cite{zhang2010collective, ramaswamy2003active, berthier2013non}, active self-organization~\cite{gokhale2021dynamic, hagan2016emergent, redner2013structure, mishra2010fluctuations},  and novel rheology~\cite{hatwalne2004rheology,saintillan2018rheology, menon2010active, haines2008effective, gyrya2011effective}; none of which are observed in systems at thermodynamic equilibrium.

To obtain an understanding of the bulk properties of active baths, investigations are often focused on length scales much larger than the individual active particles using techniques such as microviscometers or macroscopic rheometers~\cite{saintillan2018rheology}.  These studies were the first to reveal the intriguing observation of superfluidity in suspensions of swimming bacteria~\cite{sokolov2009reduction, gachelin2013non, lopez2015turning}.  This superfluid-like behavior results from a macroscopic balance between viscous dissipation and the input energy of the swimming bacteria~\cite{saintillan2018rheology}, and have motivated a large number of theoretical studies~\cite{burkholder2020nonlinear, burkholder2019fluctuation, takatori2017superfluid, saintillan2010dilute, maes2020fluctuating, ye2020active, liu2021viscoelastic, puglisi2017clausius, chaki2019effects, chaki2018entropy, lau2009fluctuating, knevzevic2021oscillatory}.  Some studies have investigated the local dynamics at the microscopic scale and how they might give rise to novel bath properties~\cite{guo2018symmetric, burkholder2019fluctuation,burkholder2020nonlinear, liu2019rheology, takatori2017superfluid,gokhale2021dynamic}. And at the scale of individual swimmers, investigations have revealed complex dynamics that depend on the local environment~\cite{mathijssen2019oscillatory,figueroa2015living,junot2019swimming,figueroa20203d, patteson2015running, ipina2019bacteria, martinez2021active}, which could play a role in the bulk active bath properties.  Pioneering studies at the microscopic scale showed: enhanced tracer diffusion~\cite{wu2000particle, jepson2013enhanced, gregoire2001active} and force fluctuations~\cite{soni2003single}, power-law stress fluctuations and violation of FDT~\cite{chen2007fluctuations}, and a memory-less friction kernel~\cite{maggi2017memory}.

Here, we use a well-established model system for creating a microscopic active bath --- a suspension of swimming \emph{E. coli}~\cite{schwarz2016escherichia, berg2008coli} --- and study the enhanced local dynamics of an immersed probe particle due to active fluctuations.  We employ approaches from non-equilibrium statistical mechanics~\cite{ahmed2015active, harada2005equality, chen2007fluctuations} and colloid physics~\cite{wilson2011microrheology,wilson2011small,squires2005simple} to measure the local fluctuations and rheology at the colloidal scale. Building on previous work~\cite{chen2007fluctuations, maggi2017memory,soni2003single}, we study force fluctuations by direct measurement of a passive colloidal probe in an active bath using an optical trap and the photon momentum method (PMM). We use a moderately dense concentration ($\phi_\mathrm{eff} = 0.2$) to create an isotropic active bath, where no long range structures, flows, or orientational order are observed and test recent theoretical predictions for the non-equilibrium  properties~\cite{burkholder2020nonlinear}.  Specifically we investigate the effect of the active bath on the local fluctuations, microrheology, and relaxation of an immersed probe particle in a regime that has not yet been explored: moderate density and nonlinear response.  

We find at Pe = 0 the probe experiences enhanced force fluctuations and the active bath approaches the solvent viscosity, but not below; at intermediate Pe (0.85 to 5.1) the active bath shear thickens to $\sim 3-5$X the viscosity of a comparable passive colloidal suspension; and at high Pe (8.5 to 50.9) the effective viscosity decreases and exhibits a plateau.  Further, the amplitude of force fluctuations in the active bath depend on Pe, a behavior that is uniquely non-equilibrium.

\section{Materials and Methods}

\subsection{Sample Preparation}
\emph{Escherichia coli} are a well-characterized model system for use as active colloids~\cite{schwarz2016escherichia}.  \emph{E. coli} were purchased from Carolina Scientific (item \#155068) and used within 48 hrs of arrival. We determined the cell density in our experiments by using optical density spectrophotometry. The measured OD600 = 0.8 corresponds to $n \sim 10^9$ cells/mL, which is equivalent to an effective volume fraction of $\phi_{\mathrm{eff}}= 0.2$.  This effective volume fraction accounts for cell body and flagella bundle. Randomly oriented cells reach ``overlap" at $\phi_{\mathrm{eff}} \sim 0.1$~\cite{schwarz2016escherichia}. We sandwiched a 20 $\mu$L droplet of solution containing \emph{E. coli} in a sample chamber made from a glass slide and a coverslip (Fisher Scientific, 12-545F) with vacuum grease (Dow Corning, Z273554) to seal the chamber. Throughout the paper, this active suspension of swimming \emph{E. coli} is called the `active bath,' and `passive bath' refers to a sample of only water. The inset of Fig.~\ref{fig:fig1} shows a representative image. Measurements were made in the middle of a sample chamber of height $\sim$300 $\mu$m and thus hydrodynamic effects due to the confined geometry have a small effect on measured viscosity ($<2\%$), as estimated by Faxen's correction for a microsphere near a boundary~\cite{leach2009comparison}.

\subsection{Optical Tweezer Measurements}

For microscopy and optical trapping, we used a Nikon TE2000 with a 60x/1.2NA water-immersion objective and Hamamatsu ORCA-Flash4.0 V2. The optical tweezer system (Impetux Optics S.L.) includes the optical trap, piezo stage positioning, and force detection. The 60x objective focuses the near-infrared laser (1064 nm, IPG-YLR-10, IPG Photonics) to create the optical trap. The photon momentum method (PMM)~\cite{farre2010force, gieseler2020optical} was implemented with a 1.4NA oil immersion condenser and a position sensitive sensor, digitized at 50 kHz, allows for force detection and laser tracking interferometry. For the force calibration to be accurate, it is critical to use a condensing objective with higher numerical aperture than the trapping objective and to minimize scattering of light through the sample~\cite{farre2010force,jun2014calibration}. The PMM approach provides direct access to the optical trapping force, even in the nonlinear regime, and does not depend on linear calibration of position and trap stiffness. A drawback of this approach is we do not have direct access to position information for trajectory analysis. Labview (National Instruments) was used to control all experimental hardware and data acquisition. 

\begin{figure}[th]
    \includegraphics[width = 0.5 \textwidth]{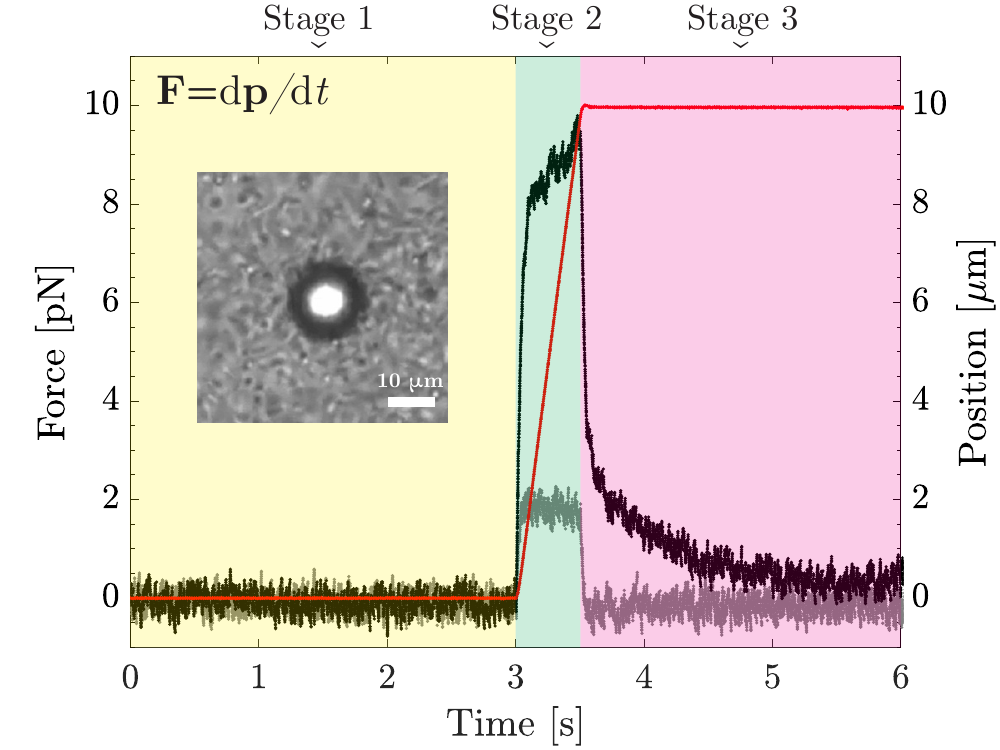}
    \caption{ Overview of experimental protocol: Spontaneous force fluctuations are measured in stage 1 (left, yellow shaded), nonlinear mechanical response in stage 2 (middle, green shaded), and force relaxation in stage 3 (right, magenta shaded).  The left vertical axis shows the optical trap force, measured via PMM, in the direction parallel to stage motion for an active (black) and passive (grey) bath. The right vertical axis shows the stage position (red), indicating time periods of no motion (stage 1 and 3) and constant velocity (stage 2).  Data shown for constant velocity of $\langle \mathbf{U} \rangle = 20$  $\mu$m/s (Pe = 8.5).  Already evident in stage 2 (green shaded) is the increased viscosity of the active suspension relative to water and the viscoelastic-like relaxation in stage 3 (magenta shaded). Left inset show representative image of a colloid optically trapped in the active bath (scale bar = 10 $\mu$m).  Note that duration of stage 1 and 3 are much longer, but not shown for clarity.}
    \label{fig:fig1}
\end{figure}

We used a colloidal probe, $r = 5$ $\mu$m,(Alfa Aesar, 42717) as our passive tracer particle for all optical tweezer measurements. We chose a probe size that allowed us to measure length scales larger than individual \emph{E. coli}. Force measurements were conducted separately on both active and passive baths. There are three distinct stages in each measurement as shown in Fig.~\ref{fig:fig1}: stage 1, the spontaneous force fluctuations of the probe (piezo is stationary); stage 2, nonlinear microrheology of the probe moving through the bath at constant velocity, $\langle \mathbf{U} \rangle,$ covering a range from 2--120 $\mu$m/s; and stage 3, force relaxation of the probe as it recovers from stage 2 perturbation back to its equilibrium position.  A representative example experiment is shown in Fig.~\ref{fig:fig1}.

\subsection{Data Analysis}
\begin{figure*}[ht]
    \includegraphics[width = 1\textwidth]{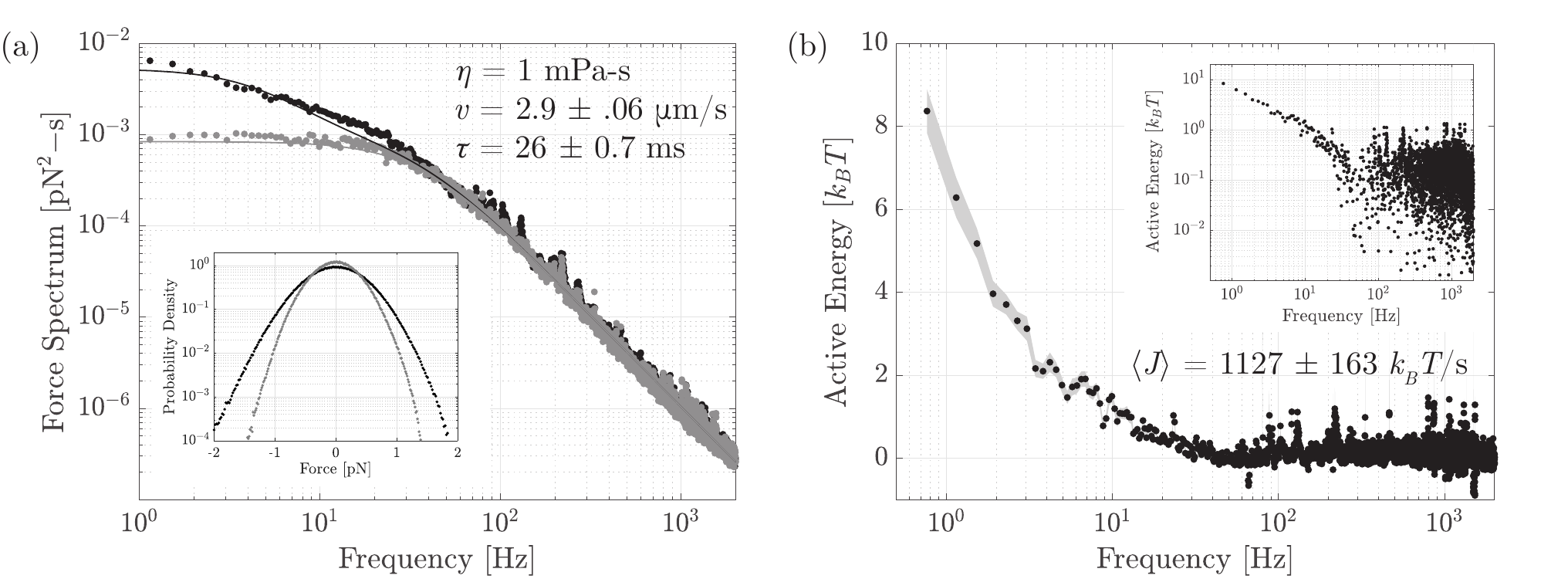}
    \caption{ Non-equilibrium fluctuations in an active bath:  (a) Force spectra calculated from stage 1 are shown for a colloid in an active (black) and passive bath (grey).  At low frequencies ($f < 50$ Hz) the force spectra in the active bath is greater.  Theoretical fit to equation~\ref{eqn:psd} (solid lines) estimates the viscosity of the bath ($\eta = 1$ mPa-s), and characterizes the active process on average by it's active burst velocity ($v = 2.9 \pm 0.06$ $\mu$m/s) and timescale ($\tau = 26 \pm 0.7$ ms).  Fit for the passive bath (solid grey) corresponds the same viscosity as the active bath ($\eta = 1$ mPa-s) and zero activity ($v,\tau = 0$).  Inset shows the probability density of force fluctuations for an active (black) and passive (grey) bath. (b) The active energy spectrum quantifies the non-thermal energetic fluctuations of a colloid in the active bath (shaded grey region indicates standard error of the mean). Integrating this spectrum provides an estimate of the energy dissipation rate, $\langle J \rangle$, via the Harada-Sasa equality~\cite{harada2005equality}.  Inset shows log-log plot of active energy spectrum.}
     \label{fig:fig2}
\end{figure*}

We calculate the force spectrum $\langle \vert \tilde{\mathbf{F}} \vert^2 \rangle $ (Fig.~\ref{fig:fig2}a), by estimating the power spectrum of a finite force signal, $\mathbf{F}(t)$, sampled at 50 kHz, using Welch's method with a Hamming window~\cite{welch1967use}.  We fit the force spectrum to our analytic model using nonlinear least squares~\cite{coleman1996interior}. 
The active energy spectrum (Fig.~\ref{fig:fig2}b) was calculated by taking the ratio of force spectra from the active and passive bath and subtracting thermal (passive) fluctuations, $E_{\mathrm{act}} = \langle \vert \tilde{\mathbf{F}}_{\mathrm{active}} \vert^2 \rangle / \langle \vert \tilde{\mathbf{F}}_{\mathrm{passive}} \vert^2 \rangle -1$, where $E_{\mathrm{act}}$ is a function of frequency in units of $k_B T$~\cite{jones2021stochastic}.  All probability distributions were calculated assuming stationarity and normalized such that, $\int  \mathcal{P}(F) dF = 1$, where $\mathcal{P}$ is the probability density, and $F$ is the force of interest. For example, in the inset of Fig. \ref{fig:fig2}(a) we use the forces from stage 1 while in Fig. \ref{fig:fig4} we use the force measured during stage 2. All data analysis was completed in MATLAB.

\subsection{Microrheology}

To measure the nonlinear response we use Pulling Active Microrheology (PAM)~\cite{wilson2011microrheology,wilson2011small,robertson2018optical}, where the probe is pulled through the sample at constant velocity for a duration of $0.5-1$ s.   While these types of measurements are often ``mixed mode'', as in neither constant force nor velocity, our experimental parameters (i.e.~stage velocity, probe size, bath particle size) put us well within the regime for constant velocity~\cite{squires2005simple}.  In PAM, it is common to define a `generalized Stokes relation', which relates the average force $\langle \mathbf{F} \rangle$, taken from the finite force signal, on the probe particle to its average velocity $\langle \mathbf{U} \rangle$, where $\langle \mathbf{F} \rangle = 6 \pi r \eta_{\mathrm{eff}} \langle \mathbf{U} \rangle$, $\eta_{\mathrm{eff}}$ is the effective viscosity of the suspension, and $\mathbf{U}$ is the velocity of the stage, ranging from $2-120$ $\mu$m/s.  We use this relation to calculate the effective viscosity of our suspension as a function of P\'{e}clet number (Pe). Pe is defined for active suspensions as, $\mathrm{Pe} = \dot{\gamma} \tau_r$~\cite{burkholder2020nonlinear}, where $\dot{\gamma} = 3\langle \vert \mathbf{U} \vert \rangle / \sqrt{2} r$ is the shear rate and $\tau_r$ is the persistence time of the active bath particle.  For \emph{E. coli}, persistence times are roughly one second~\cite{patteson2015running, figueroa20203d}, so we use $\tau_r = 1$ s for simplicity, such that $\mathrm{Pe} = \dot{\gamma}$.  Therefore in our PAM experiments we explore the regime, $0.85 < \mathrm{Pe} < 50.9$.  This data is color coded in figures as Pe = 0.85 (brown), 1.7 (red), 3.4 (orange), 5.1 (yellow), 8.5 (green), 17.0 (cyan), 33.9 (blue), 50.9 (royal).  We refer to this as nonlinear rheology for two reasons: (1) the forces measured (via PMM) are  outside the linear regime of the optical trap and (2) the measured viscosity has a nonlinear relationship to shear rate.

To estimate the effective viscosity of our moderate volume fraction suspension in the absence of activity, we employ the widely used Krieger-Dougherty relation, $\eta_{\mathrm{eff}} / \eta_0 = (1 -\phi / \phi_{max})^{-2}$ ~\cite{krieger1959mechanism,quemada1977rheology,ball1980dynamics,van1989hard, boek1997simulating}, where $\phi$ is the volume fraction and $\eta_0$ is the viscosity of the background solvent.  Using our volume fraction $\phi = \phi_{\mathrm{eff}} = 0.2$ and a $\phi_{max} = 0.63$ (for spherical packing) we estimate the viscosity of an equivalent isotropic passive suspension to be $\eta_{\mathrm{eff}} / \eta_0 = 2.15$. This value provides a baseline expected viscosity of  suspension without activity.

\subsection{Theoretical model}

As discussed previously~\cite{jones2021stochastic}, we model the stochastic forces of the optically trapped colloidal probe subject to thermal and nonthermal forces with the overdamped Langevin equation~\cite{fodor2018statistical,sekimoto1998langevin}.  That is, the position $\mathbf{r}(t) \in \mathbb{R}^2$ of the colloidal particle is governed by
\begin{equation} 
    \gamma\dot{\mathbf{r}}+\kappa \mathbf{r}=\gamma \mathbf{u}+\sqrt{2D}\gamma \boldsymbol\xi,
    \label{eqn:EoM}
\end{equation}
which balances the deterministic frictional and optical trap forces with the random active and thermal forces. In equation \eqref{eqn:EoM}, $\kappa$ is the optical trap stiffness, $\gamma = 6 \pi R \eta$ is the friction coefficient of the Stokes' drag where $R$ is the particle radius and $\eta$ is viscosity, and $D$ is the thermal diffusion coefficient of the zero-mean, $\delta$-correlated Gaussian white noise process $\boldsymbol\xi$. In this model, active bath particles collide with the probe particle, transferring energy, which manifests in an active burst velocity of the probe, $\mathbf{u}$.  This active burst velocity, $\mathbf{u}$, is modeled as an Active Ornstein-Uhlenbeck (AOUP) process with characteristic strength, $v$, timescale, $\tau$, and correlations $\langle u(t) u(s) \rangle = v^2 \exp ^{-\vert t - s \vert / \tau} $~\cite{jones2021stochastic, fodor2018statistical, martin2021statistical}.  This model allows for an analytic form of the force spectrum, as derived previously~\cite{jones2021stochastic},

\begin{equation}
\begin{split}
S\!_{f\!f}(\omega) = & \left(2\kappa^{2}D-\frac{2\tau\kappa^{2}v^{2}}{\mu^{2}\tau^{2}-1}\right)
\frac{1}{\mu^{2}+\omega^{2}}
\\ & +\frac{2\tau\kappa^{2}v^{2}}{(\mu^{2}\tau^{2}-1)}
\frac{1}{\tau^{-2}+\omega^{2}},
\end{split}
\label{eqn:psd}
\end{equation}
where $\omega$ is frequency in rad/s, $\mu = \kappa/\gamma$, and $v = \vert \mathbf{u} \vert$. Equation \eqref{eqn:psd} is fit to the experimentally measured force spectrum measured in stage 1 to extract non-equilibrium activity.  We note that more complex and realistic models of the active process could be used specifically for \emph{E. coli}~\cite{figueroa20203d, baskaran2009statistical}, however, we use the AOUP process for generalizability to non-bacterial active baths and analytic tractability~\cite{martin2021statistical}.

\section{Results and Discussion}
\subsection{Stage 1: Non-equilibrium force fluctuations}

To extract the non-equilibrium force fluctuations on the colloid due to the active bath we focus on stage 1 (Fig.~\ref{fig:fig1}, yellow) where the colloid is fluctuating in the stationary trap due to forces from the surrounding media.  We use the force spectrum, a commonly used approach~\cite{ahmed2015active}, to quantify the force fluctuations on a colloid in both an active and passive bath. The average force spectra, shown in Fig.~\ref{fig:fig2}a for active (black dots) and passive (grey dots) baths, exhibit two notable features: (1) In the high frequency regime ($f \gtrsim 50$ Hz) they collapse on one another; and  (2) at lower frequencies ($f \lesssim 50$ Hz) the two curves diverge.  These measurements show that in the active bath the high frequency fluctuations are dominated by thermal forces and the low frequency fluctuations are dominated by non-thermal forces due to activity, as seen previously~\cite{chen2007fluctuations, maggi2017memory, bohec2019distribution}.  

By fitting the analytic equation for the force spectra (equation~\ref{eqn:psd}) we extract several physical parameters from the model, \textcolor{blue}{specifically, $\eta$, $v$, and $\tau$}. Interestingly, the extracted viscosity was indistinguishable between the active and passive bath, $\eta = 1$ mPa-s, during stage 1.  This suggests that the swimming bacteria in the active bath do not contribute to the overall suspension viscosity as experienced by the colloid but do contribute to enhanced fluctuations~\cite{hatwalne2004rheology,haines2008effective,saintillan2010dilute, lopez2015turning,guo2018symmetric, burkholder2019fluctuation,burkholder2020nonlinear}.  This dichotomy highlights the lack of a direct connection between fluctuation and dissipation. 

The two parameters that characterize the microscopic activity of the bath are the burst velocity and timescale (which are both zero for a passive bath).  The burst velocity, $\langle v \rangle = 2.9 \pm 0.06$ $\mu$m/s, and timescale, $\langle \tau \rangle = 26 \pm 0.7$ ms, represent the average non-equilibrium fluctuation transmitted to the probe colloid from the active bath.  Both of these values are smaller than those of a single swimmer~\cite{figueroa20203d}, as expected since the probe colloid is larger and its motion is likely due to many collisions.  The burst velocity allows estimation of the average non-equilibrium force fluctuation to be approximately, $F = 6 \pi r \eta v \approx 0.3$ pN, sustained for an average time $\tau$.  This activity is manifested in a wider distribution of force fluctuations experienced by the probe during stage 1 in the active vs passive bath (Fig.~\ref{fig:fig2}, inset).  This type of non-Gaussian force fluctuations are common in non-equilibrium systems~\cite{leptos2009dynamics,rushkin2010fluid}.

To further characterize the non-equilibrium fluctuations we calculate the active energy spectrum from the ratio of the force spectra~\cite{jones2021stochastic, eldeen2020quantifying}. Fig.~\ref{fig:fig2}b shows that low-frequency active fluctuations have energy scales on the order of $k_B T$, which corresponds to a dissipation rate of $\langle J \rangle \approx 10^3$ $k_B T/$s when integrated over all available frequencies,  \emph{a la} the Harada-Sasa equality~\cite{harada2005equality,jones2021stochastic}.  This value, $\langle J \rangle $, estimates the average rate of energy transferred from the active bath to the probe colloid that is manifested in translational fluctuations.  It is worth noting, that this value is remarkably close to the power dissipated by an individual swimming bacteria~\cite{ishikawa2011energy, chattopadhyay2006swimming}, but this is likely a coincidence since the overall motion of the probe colloid is presumably due to a large number of collisions. To quantitatively relate the average burst velocity/timescale, force fluctuation, and dissipation rate of the active bath to the individual swimming bacteria requires a detailed micromechanical model that considers momentum exchange --- a topic of future work.

Overall, analysis of spontaneous force fluctuations during stage 1 (stationary piezo stage) allows characterization of the amplitude, timescale, and energetics of non-equilibrium fluctuations experienced by the probe colloid in an active bath.  The parameters extracted from our theoretical fit provides an estimate of the effective self-diffusion of the probe colloid in an active bath where, $D_{\mathrm{eff}} = D_{\mathrm{thermal}} + D_{\mathrm{active}}$, where $D_{\mathrm{thermal}} = k_B T / 6 \pi r \eta$ and $D_{\mathrm{active}} = v^2 \tau / 6$~\cite{takatori2014swim}.  We find that the thermal and active diffusion coefficients are $4.4 \times 10^{-14}$ m$^2$/s and $3.6 \times 10^{-14}$ m$^2$/s, respectively, indicating that the active bath almost doubles the effective diffusion of the probe colloid at long timescales.  

\begin{figure*}[ht]
    \includegraphics[width = 1 \textwidth]{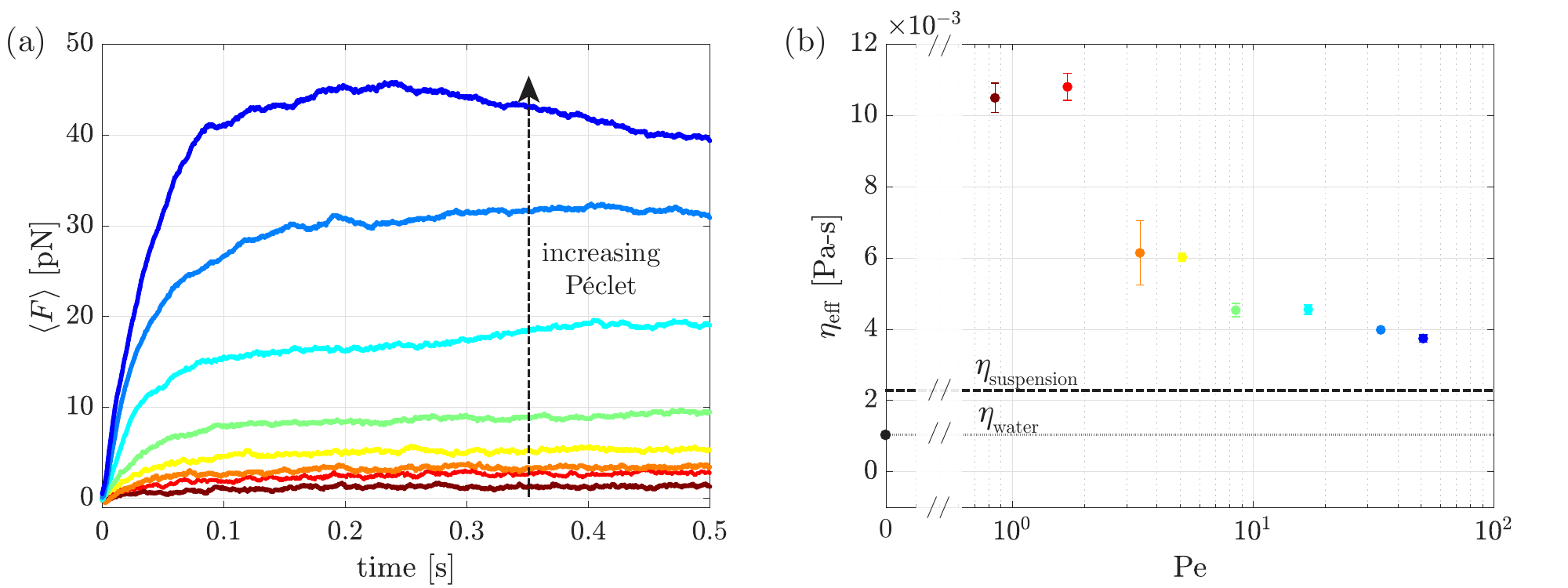}
    \caption{Nonlinear microrheology and effective viscosity: (a) Ensemble averaged force as a function of time, $\langle \vert \mathbf{F}(t) \vert \rangle$, is shown for measurements in stage 2 (constant velocity) for Pe = 0.85 (brown), 1.7 (red), 3.4 (orange), 5.1 (yellow), 8.5 (green), 17.0 (cyan), 33.9 (blue), 50.9 (royal).  Plateau forces clearly increase with Pe. (b) Effective viscosity, $\eta_{\mathrm{eff}}$, as a function of Pe.  Dashed horizontal lines show the viscosity of water ($\eta_{\mathrm{water}}$) and expected viscosity of the passive suspension ($\eta_{\mathrm{suspension}}$) according to the Krieger-Dougherty relation.  Note the horizontal axis break (//) to allow a data point at zero Pe, where the active bath has an effective viscosity equivalent to water.  Increasing Pe causes shear thickening with decreasing strength.
    }
    \label{fig:fig3}
\end{figure*}

\subsection{Stage 2: Nonlinear microrheology and effective viscosity}

To characterize the response of the active bath to an applied force we focus on stage 2 (Fig.~\ref{fig:fig1}, green) where the colloid is pulled through the active suspension at constant velocity (or Pe).  The force response increases in magnitude with increasing Pe, and shows an initial slope followed by a plateau, as shown in Fig.~\ref{fig:fig3}a.  To quantify the viscous response of the active bath we calculate the effective viscosity ($\eta_{\mathrm{eff}}$, see methods) which exhibits a strong dependence on Pe (Fig.~\ref{fig:fig3}b).  

A recent theoretical study has highlighted that the effective viscosity is strongly dependent on local length scales~\cite{burkholder2020nonlinear}, but experimental comparisons are lacking.  For direct comparison to this recent theoretical work we introduce two length scales: The distance moved by the colloidal probe during a characteristic time, $L = U\tau_L$, where $U = \langle \vert \mathbf{U} \vert \rangle$ is the probe speed maintained for a time $\tau_L$.  The distance moved by the active bath particle, $\ell = v \tau_r$, where $v$ is the self-propulsion speed and $\tau_r$ is the persistence time.  Here, since both timescales are of $\mathcal{O}(1)$, then $L/\ell \sim U/v$.  Further we introduce the center-to-center separation distance of the probe and active particle upon contact to be $R_c = r + a$ where $r$ is the size of the colloidal probe and $a$ is the size of the active particle.  In our experiments $\ell / R_c \sim 1$, meaning the distance traveled by an active bath particle during its characteristic reorientation time, $\tau_r$, is comparable the center-to-center distance of the probe and active particle pair.

At ``zero" Pe (or zero shear rate), where $L \ll \ell$, the viscosity is equivalent to that of water even though it is actually a dense suspension. This zero shear viscosity is deduced from the stage 1 fit, because $\eta_{\mathrm{eff}}$ is not defined for $\mathbf{U} = 0$.  In this regime, the active bath particles travel a much greater distance than the probe during a time, $\tau_r$, and are able to transfer force to the probe from all directions. According to the Krieger-Dougherty relation, a passive suspension of $\phi_{\mathrm{eff}}=0.2$ the expected viscosity is $\eta_{\mathrm{eff}} \approx 2$ mPa-s, whereas our measured viscosity is roughly half that.  This is consistent with several previous studies that have found superfluid behavior of active suspensions due to an effective shear thinning caused by the active swimmers~\cite{takatori2017superfluid, burkholder2020nonlinear, saintillan2018rheology, lopez2015turning, gachelin2013non, guo2018symmetric, chui2021rheology}.  However, our data indicates that the effective microviscosity decreases to the solvent viscosity but we do not observe further thinning.  This discrepancy between previous studies could be due to measurements that probe different length scales and/or the absence of large scale ordering in our active bath.  Specifically, the microviscosity measured here characterizes the local environment at the colloidal scale ($r = 5$ $\mu$m), whereas previous measurements were of bulk environments with length scales on the order of $10^2$ $\mu$m for microfluidic viscometers~\cite{gachelin2013non, liu2019rheology} or $10^3 - 10^4$ $\mu$m for macrorheometry~\cite{lopez2015turning, chui2021rheology}. Our measured thinning of the active bath down to the solvent viscosity, but not lower, is in agreement with recent simulations of an isotropic active bath~\cite{burkholder2020nonlinear}.  Superfluidization to levels below the solvent viscosity may require large-scale shear to organize flow fields~\cite{guo2018symmetric, sokolov2009reduction} and may only be evident on larger lengthscales.

At intermediate Pe (0.85 to 5.1), where $L/ \ell \sim 1$, we see a large increase in effective viscosity of three to five times the value expected for a passive suspension.  In this regime, the probe particle and the active bath particles move comparable distances during a time, $\tau_r$. The proposed explanation~\cite{burkholder2020nonlinear} is that active particles behind the probe have difficulty pushing because they are moving at roughly the same speed, whereas the opposite is true for active particles in front of the probe that are able to push backward on the probe --- leading to force thickening.  It is interesting to note that this mechanism for force thickening~\cite{burkholder2020nonlinear} is completely independent of hydrodynamic lubrication interactions as occurs in passive colloidal suspensions~\cite{cheng2011imaging, pednekar2017simulation}.  Recent simulations further support the above mechanism of force thickening, namely an inhomogeneous distribution of active particles~\cite{knezevic2021oscillatory}.

At large Pe (8.5 to 50.9), where $L > \ell$, we see a relative decrease in the effective viscosity (or a decreasing effect of shear thickening) that seems to plateau.  In this regime, the probe particle moves much further than the active bath particle during the reorientation time $\tau_r$.  Here, in line with Burkholder and Brady~\cite{burkholder2020nonlinear}, we expect that the active particles behind the probe are not able to fill in the wake left by the probe motion and active particles in front are not able to escape and accumulate in the boundary layer.  Essentially, at high probe velocities the active bath cannot ``keep up" and the effective viscosity resembles that of a passive suspension exhibiting a high Pe plateau.  The measured plateau is roughly twice the expected viscosity for a passive suspension of hard colloidal spheres  estimated via the Krieger-Dougherty relation for $\phi_{\mathrm{eff}}=0.2$~\cite{krieger1959mechanism}.  This larger plateau viscosity could be due to non-spherical geometry or interactions between the active bath particles (\emph{E.~coli}), which are unaccounted for in this estimate.

Overall, our nonlinear microrheology results largely confirm theoretical predictions~\cite{burkholder2020nonlinear}: At  Pe $\ll 1$, the active bath particle motion dominates and thins the suspension leading to a reduced zero-shear plateau viscosity equivalent to that of the solvent (water).  At intermediate Pe, where the motion of the probe particle and active bath particle are comparable ($L \sim \ell$), the active bath particles push backward on the advancing probe and the suspension thickens.  At Pe $>8.5$,  the probe motion dominates ($L > \ell$) and the active bath particles cannot keep up with its motion leading to a plateau viscosity as seen in passive suspensions.  This type of non monotonic shear thickening at the single particle level is qualitatively similar to previous theoretical predictions in dilute suspensions~\cite{saintillan2010dilute}.

\begin{figure*}[th]
    \includegraphics[width = 1 \textwidth]{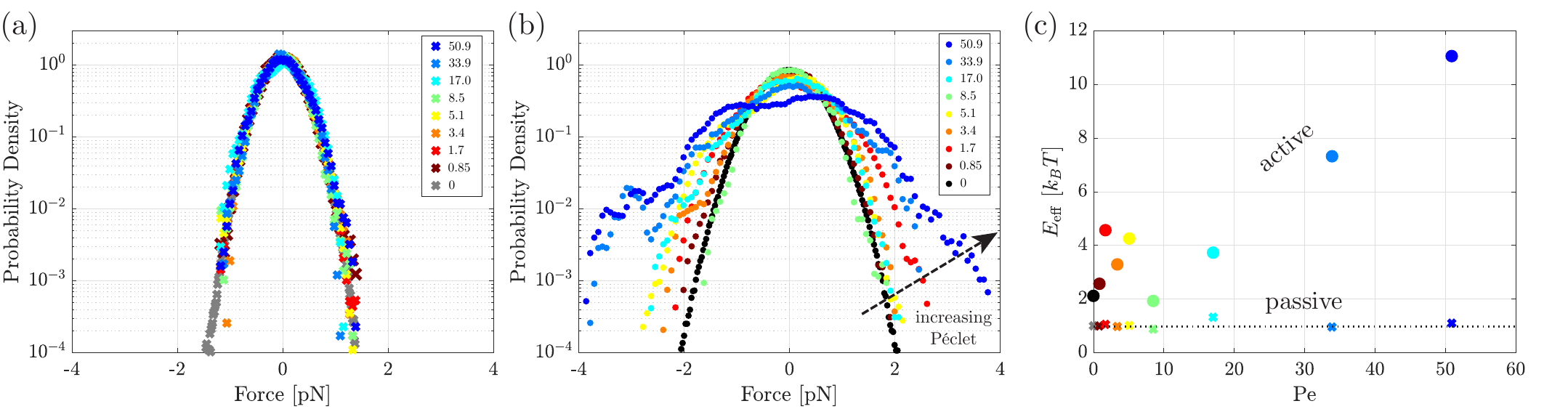}
    \caption{Force fluctuations depend on P\'{e}clet number: Colored symbols in legend indicate Pe for each data set. (a) In a passive bath, the probability density of force fluctuations shows no dependence on Pe. (b) In an active bath, the distribution of forces clearly widens with increasing Pe. (c) The effective energy of the active bath consistently exhibits fluctuations greater than equilibrium and this effect increases with Pe. Dashed horizontal line indicates thermal equilibrium, $E_\mathrm{eff} = k_B T$, (symbols differentiate data for the passive ($\times$) and active bath ($\bullet$)).  }
    \label{fig:fig4}
\end{figure*}

\subsection{Force fluctuations depend on P\'{e}clet number}

An advantage of measuring the nonlinear response of the active bath using optical tweezer microrheology is access to full information on the force fluctuations experienced by the probe~\cite{wilson2009passive}.  These fluctuations are related to effective temperature relations and non-equilibrium work theorems~\cite{kurchan2007non,wilson2011microrheology, evans2002fluctuation}.  To focus on the fluctuations, we analyze the direction orthogonal to PAM to remove the direct influence of the trap motion and plot the probability density of force fluctuations in Fig.~\ref{fig:fig4}.  For the passive bath composed of a Newtonian solvent (i.e water) the force fluctuations orthogonal to trap motion do not depend on Pe (Fig.~\ref{fig:fig4}a).  

For the active bath suspension, the fluctuations increase with Pe and become non-Gaussian (Fig.~\ref{fig:fig4}b). We can characterize this by plotting the effective energy, $E_\mathrm{eff}$, of the particle as a function of Pe (Fig.~\ref{fig:fig4}c).  We calculate $E_\mathrm{eff}$ from the variance of the force distributions in Fig.~\ref{fig:fig4}a,b;  assuming the fluctuations in the passive bath at Pe = 0 have an energy of $k_B T$ (grey, $\times$).  Specifically, the effective energy was calculated as $E_\mathrm{eff} = \sigma^2 / \sigma^2_{\mathrm{grey,} \times}$, where $\sigma^2$ is the variance of the force distribution of interest and $\sigma^2_{\mathrm{grey,} \times} $ is the variance of the force distribution at Pe = 0 in a passive bath.

The $E_\mathrm{eff}$ plotted in Fig.~\ref{fig:fig4}c shows a clear dependence on Pe for the active bath ($\bullet$) but not for the passive bath ($\times$).  At low Pe, this relationship is weak but the $E_\mathrm{eff}$ of the active bath is always greater than the passive bath.  As Pe increases, the $E_\mathrm{eff}$ of the active bath clearly increases.  Enhanced fluctuations at high Pe (but not low Pe) have also been observed in dense suspensions of passive colloids~\cite{wilson2011microrheology}.  Therefore, we interpret this as follows: At low Pe, enhanced fluctuations come primarily from the activity of the bath particles.  At high Pe, enhanced fluctuations are a combination of activity and steric interactions due to the probe pushing active bath particles at high shear rate.  This interpretation is consistent with our effective viscosity measurements (Fig.~\ref{fig:fig3}b) where at Pe $< 8.5$ activity of the bath plays an important role, whereas at Pe $\geq 8.5$ a viscous plateau is observed, qualitatively consistent with passive colloidal suspensions.

\begin{figure}[ht]
    \includegraphics[width = 0.5 \textwidth]{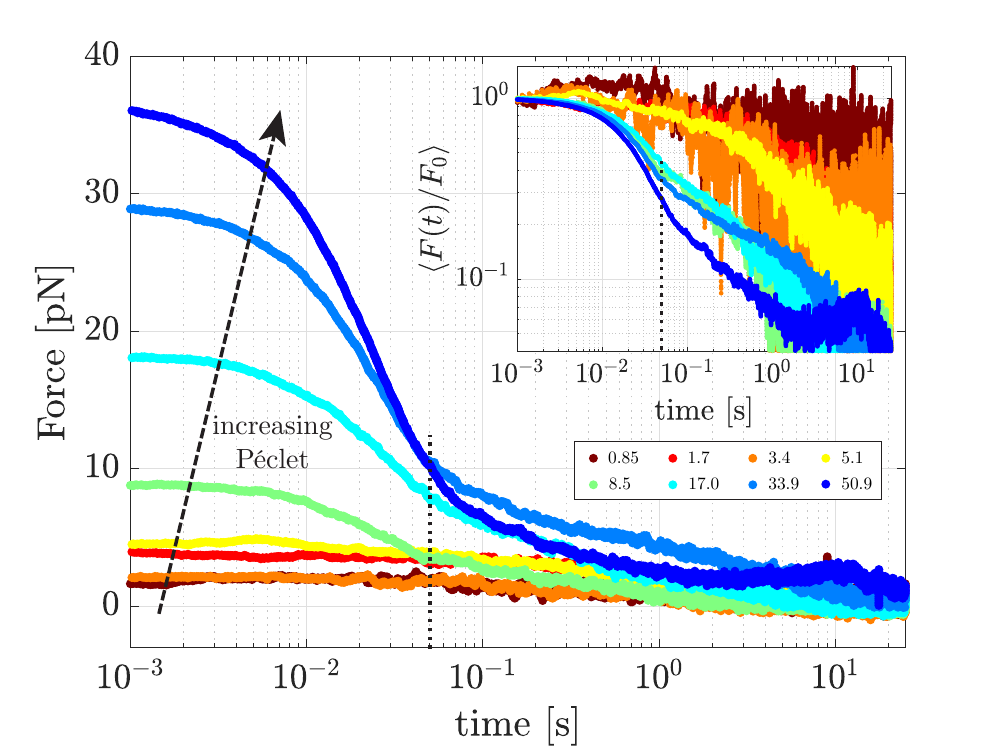}
    \caption{Force relaxation depends on Pe: For Pe $\geq 8.5$ the force relaxation exhibits a rapid decay for $t \leq 50$ ms, followed by a slow relaxation.  For Pe $< 8.5$ the initial rapid decay is not observed. Colored symbols in legend indicate Pe for each data set.  Inset shows the normalized force relaxation.}
    \label{fig:fig5}
\end{figure}

\subsection{Stage 3: Force relaxation}

Force relaxation is observed in stage 3 (Fig.~\ref{fig:fig1}, magenta) where the optical trap is stationary and the probe is relaxing from its perturbed state in stage 2 (Fig.~\ref{fig:fig1}, green).  This force relaxation is challenging to interpret due to the large number of physical processes occurring simultaneously,~e.g.~active fluctuations, heterogeneous bath density, colloidal suspension dynamics, viscoelastic effects.  However, one clear observation is the dependence on Pe as shown in Fig.~\ref{fig:fig5}.  For Pe $ \geq 8.5$, forces exhibit an initial rapid decay during the first 50 ms (dashed vertical line in Fig.~\ref{fig:fig5}) followed by a slow relaxation to the equilibrium position.  For Pe $< 8.5$, forces exhibit only the slow relaxation to the equilibrium position.  This is in stark contrast to the nearly instantaneous relaxation for a Newtonian fluid in thermal equilibrium (Fig.~\ref{fig:fig1}, grey).  

Fig.~\ref{fig:fig5} inset is the normalized force relaxation, which accentuates this effect showing a rapid decay for Pe $\geq 8.5$ and a slow relaxation for Pe $< 8.5$.  We interpret this as follows:  At Pe $\geq 8.5$, the rapid decay is due mainly to the passive properties of the bath, where large forces cause steric rearrangements of the bath particles.  This is followed by a slower force relaxation due to the active fluctuations at $t > 50$ ms.  This is consistent with measurements of the force spectra (Fig.~\ref{fig:fig2}a) where the timescale of the active process was estimated to be $\langle \tau \rangle = 26$ ms and thus the effects of such activity is visible on timescales greater than that.  At Pe $< 8.5$, the force relaxation is dominated by the bath activity, because steric interactions due to large deformation are absent.  Recent work suggests that activity of the bath can have a strong effect on relaxation dynamics~\cite{caprini2021fluctuation,caprini2021generalized}.  To interpret the relaxation, extension of active viscoelastic models~\cite{banerjee2021active} to the nonlinear regime and incorporating active elements into a viscoelastic memory kernel are promising next steps. These models must consider the microscopic dynamics (e.~g.~ the detailed interactions between the probe and bath particles) --- a topic reserved for future work.

\section{Conclusion}

Altogether, our results show that an immersed micron-scale probe in a moderately dense ($\phi_\mathrm{eff} = 0.2$) active bath of \emph{E. coli} experiences, on average, active forces of $\langle F \rangle \sim 0.3$ pN for a duration of $\langle \tau \rangle = 26 $ ms resulting in non-thermal energy transfer of $\langle J \rangle \approx 10^3 k_B T/$s.  This results in enhanced diffusion at long timescales and superfliud-like thinning down to the solvent viscosity at Pe $\sim 0$.  At intermediate Pe, the active bath shear thickens to 3--5X the viscosity of a comparable passive suspension, exhibits increased amplitude of force fluctuations and a slow relaxation back to equilibrium from its perturbed state.  At higher Pe, the active bath exhibits a viscous plateau of 2X the viscosity of a comparable passive suspension, shows force fluctuations that increase with Pe, and exhibits a rapid force decay followed by a slow relaxation to equilibrium from its perturbed state.  Our results complement previous experimental~\cite{chen2007fluctuations, maggi2017memory, wu2000particle} and theoretical studies~\cite{saintillan2018rheology, burkholder2020nonlinear}, and contribute to the emerging picture that when the distances traversed in a characteristic time by the immersed probe and active bath particle are small, $L / \ell \leq 1$, the active bath exhibits novel non-equilibrium properties; and when $L / \ell > 1$ that active bath behaves much like a passive colloidal suspension.  A natural important extension of this work would be to systematically vary the activity of the active bath, controlling $L / \ell$, to study the effect on active fluctuations and viscosity experienced by the embedded probe.  This could be accomplished using a heat exchanging fluid bath~\cite{krishnamurthy2016micrometre} or localized laser heating~\cite{blickle2012realization,martinez2016brownian}. 

\section*{Acknowledgements}
This material is based upon work supported by the National Science Foundation under Grant No. NSF DMR-2004566 to WWA, NSF DMR-2004417 to JLR, and NSF DMR- 2004400 to WBR. HS and MG were partially supported by the Dan Black Family Trust Fellowship.

\bibliographystyle{ieeetr}
\bibliography{bibtex}% Produces the bibliography via BibTeX.

\end{document}